\documentclass{article}
\usepackage[utf8]{inputenc}
\usepackage{PRIMEarxiv}
\usepackage{amsmath, amssymb}
\usepackage{graphicx}
\usepackage{todonotes}
\usepackage{multirow}
\usepackage{subfigure}
\title{Cobol2Vec: Learning Representation of COBOL code}
\author{Ankit Kulshrestha\\
Clemson University\\
\texttt{akulshr@clemson.edu}
\And
Vishwas Lele\\
Applied Information Sciences Inc.\\
\texttt{Vishwas.Lele@appliedis.com}}
\date{October 2020}

\begin{document}
\maketitle
\begin{abstract}
    There has been a steadily growing interest in development of novel methods to learn a representation of a given input data and subsequently using them for several downstream tasks. The field of natural language processing has seen a significant improvement in different tasks by incorporating pre-trained embeddings into their pipelines. Recently, these methods have been applied to programming languages with a view to improve developer productivity.\\ 
    
    In this paper, we present an unsupervised learning approach to encode old mainframe languages into a fixed dimensional vector space. We use COBOL as our motivating example and create a corpus and demonstrate the efficacy of our approach in a code-retrieval task on our corpus. 
\end{abstract}

\section{Introduction}

Programming languages have a fixed life-cycle and their maintenance is costly after the support for a language has ended. Moreover, it is required to convert programs written in obsolete languages into modern languages. This conversion often requires domain-level expertise which is hard to obtain and not always available. Ideally, we would like to automate this process of migrating an obsolete codebase into a modern framework. A first step towards this goal is to \emph{learn} a representation of the code in the source language and use it  to generate code-tokens in target language.  Deep learning techniques have shown to be effective at capturing structure in natural language~\cite{mikolov2013efficient} and more recently in tasks on programming languages like method classification~\cite{alon_code2vec_2019}, comment generation~\cite{feng_codebert_2020}, transpilation~\cite{lachaux_unsupervised_2020} etc. Given the statistical nature of our problem and the efficacy of deep learning techniques, we choose to incorporate deep neural networks into our solution pipeline.

Our main focus in this paper is to learn a good representation of the given source code. This means that we want to capture both \emph{semantics} and \emph{structure} of code in a single fixed low-dimensional vector in abstract space. For this paper, we solely focus on the COBOL programming language and tailor our input representation towards it's unique style. However, from the modeling perspective our technique is general enough to be applied to any programming language.

\emph{Challenges}: Learning representations from mainframe languages like COBOL has it's fair set of challenges. The first issue is the lack of availability of open source COBOL code which makes it hard to create scalable models for capturing program logic. Second, COBOL is a very verbose language with program statements closely resembling spoken sentences. For instance, an assignment from a single variable to another is written as ``\texttt{MOVE X TO Y}". This verbosity not only makes it hard to come up with compact representations of code but also introduces a lot of tokens in the vocabulary. The issue of user defined code tokens has also been observed in modern languages~\cite{alon_code2vec_2019}. Hence, coming up with a model that generalizes well to unseen ``out of vocabulary"(OOV) tokens is difficult and error prone. The third challenge, common to all modeling techniques on code is the sheer amount of \emph{variance} in the code logic since the same program can be written differently by different programmers based on style. In later sections we address these challenges in a more detailed manner for our specific application.

\subsection{\emph{Applications}}

A lot of modern software systems tend to be open source today. The source code of mature open source softwares like Linux, Django, Flask etc typically contain billions of ``code tokens". Given such an abundance of code tokens, it is natural to look towards statistical techniques that can identify patterns in the source code. In particular, machine learning methods can be applied to code to derive and exploit complex patterns across millions of files and generate useful insights for the programmer.  

One of the main tasks in setting up a machine learning system for code processing is the input representation. Since we want to capture both structure and semantics, our work introduces a novel concept of an \emph{abstract structure} representation of code that allows us to represent both semantics and structure in an efficient manner.  More practically, our technique can be applied in the following areas/applications:

\begin{itemize}
    \item \textbf{Code Retrieval}: Given a large codebase, retrieve specific code matching a given code snippet. This feature is useful for both business and personal use. In the personal use case, the feature can be used in Integrated Development Environments (IDEs) to quickly search for a similar function across a large codebase. In a business use case, the feature can be utilized to detect malicious code/security vulnerabilites in production software. 
    
    \item \textbf{Visual Mapping}: The learned embeddings from the Cobol2vec (or similar) model can be utilized to discover clusters of code that are similar in a visual manner as well. This can provide valuable insight into redundant code/refactoring needs of the deployed application.
\end{itemize}

\subsection{\emph{Contributions}}
In this paper, we explore learning a latent representation of mainframe  languages like COBOL using deep learning. Specifically, we have the following contributions in our work:

\begin{enumerate}
    \item Introduce an abstract structural representation  that aims to capture the semantics and the structure of COBOL code.
    
    \item Build a \texttt{seq2seq} autoencoder to learn a fixed low dimensional representation of a COBOL program.
    
    \item Experimentally verify the efficacy of the learned embeddings on code retrieval task.
\end{enumerate}

\section{Related Work}

Work on programming language is still in its nascent stages and there is a steadily growing body of research to use machine learning techniques in the domain of programming languages. Hindle~\emph{et al.}~\cite{hindle_naturalness_nodate} were the first to equate programming languages with natural language and show that programming languages are a very \emph{structured } form of communication. Allamanis~\emph{et al.}~\cite{allamanis_mining_2013} then extended this notion by noting that a code snippet not only serves as a medium of communication with the machine, but also between programmers. They train a trigram model showing several important results for learning about code.Tufano~\emph{et al.}~\cite{tufano_deep_2018} considered deep learning approaches for code retrieval task on three different representations of code. In most of the aforementioned works, the input programming language is Java and the different sources of representations are readily available. Moreover, it is fairly straightforward to obtain millions of lines of open source code in Java. We note that our considered case has neither the size nor the availability of readily available representation format. 

Mikolov~\emph{et al.}~\cite{mikolov2013efficient} introduced the concept of representing words in high dimensional vector space for various downstream tasks. Following the equivalence of~\cite{hindle_naturalness_nodate}, Alon~\emph{et al.}~\cite{alon_code2vec_2019} proposed the idea of vectorizing code fragments for method name prediction. Their approach treats code as a bag of root-to-leaf paths in an AST. They train a deep learning language model with attention to predict the method names. A drawback of this approach is that code is inherently sequential in structure and hence an unordered bag of words representation throws out \emph{context} from the raw input representation. 

In the recent times, BERT~\cite{devlin2019bert} based models have become popular due to their incredible performance on natural language tasks. CodeBERT~\cite{feng_codebert_2020} to produce general representations from a combination of natural language (e.g. Python's docstrings) and programming language for downstream tasks. They use masked language modeling techniques to train their version of BERT. A very promising line of work is introduced by~\cite{lachaux_unsupervised_2020} where a mapping from one programming language to another is achieved in an unsupervised manner. The results shown by the authors give a strong foundation for further studies on such bi-directional programming language models.

\section{Theory}

In this section we present our \emph{abstract structure} representation and compare it with existing methods of representing input programs. We then present our architectural design for learning latent representation of code.

\begin{figure}
    \centering
    \includegraphics[width=.65\columnwidth]{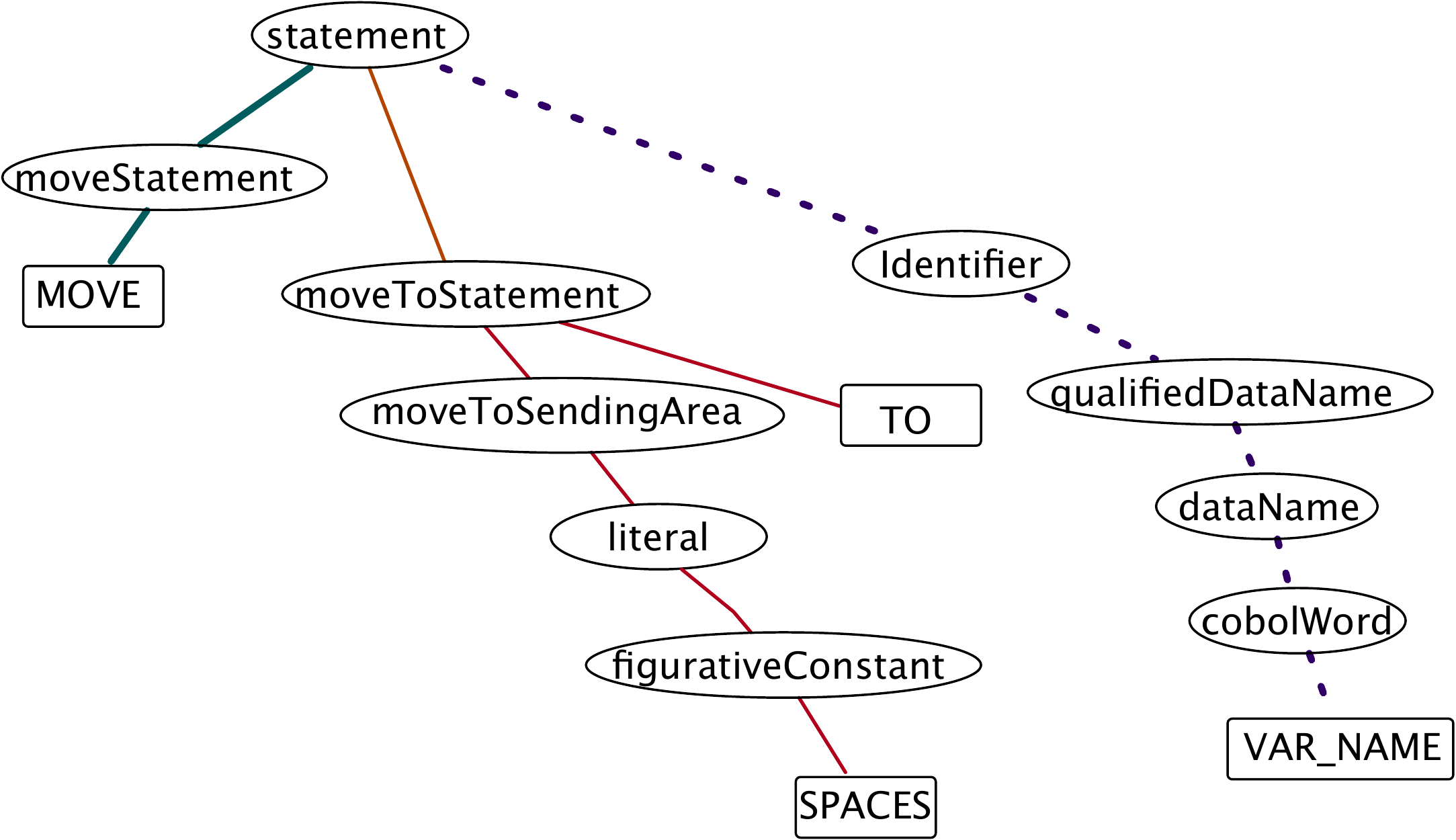}
    \caption{AST representation of a simple COBOL snippet. The square boxes represent the terminal nodes and the oval boxes represent internal nodes of the AST.}
    \label{fig:ast_example}
\end{figure}

\subsection{Why is AST inefficient for COBOL?}

Modern programming languages like Python and Java are symbolic languages i.e. they have a set of operators and they act upon different variables introduced in a particular Java method or Python script. In these languages, a program's structure can be understood by analyzing the AST representation of code that shows the heirarchical relationship between an abstract operator and its operands. By flattening out root to leaf paths in an AST and vectorizing them~\cite{alon_code2vec_2019}, one can learn a structural relationship between different statements in the program, assuming the order of statements is never changed across the codebase.

 COBOL on the other hand, does not have support for symbolic computation. All statements are verbose instructions comprising of a keyword followed by operands in a manner similar to how one would \emph{explain} a symbolic code verbally. Moreover, the code is organized into an atomic unit via paragraphs that are simply strings with a period at their end. Analyzing such verbose languages via AST is not only cumbersome but also does not yield a lot of insight since statements in a COBOL paragraph are not necessarily locally correlated.  An example of an AST of a simple ``MOVE" statement belonging to some paragraph is shown in Figure~\ref{fig:ast_example}. It is evident that if we persist with flattening the root to leaf paths for a COBOL paragraph, the number of spuriously correlated vectors will drive the quality of any machine learning algorithm downwards.

\subsection{Abstract Structure Represenation}
Let $\Lambda$ be set of all programming language specific constructs. For instance, all COBOL statements like \texttt{PERFORM}, \texttt{ADD}, \texttt{MOVE} belong to this set. Further, let $\Gamma$ be the set of all user defined variables and $\mathbb{I}$ be the set of all COBOL identifiers in program. We also define $\phi: \Gamma \rightarrow \tau_{u}$ to be a mapping function that maps the user defined variables to a single special (abstract) token and $\psi: \mathbb{I} \rightarrow \tau_{I}$ to be another mapping function that  maps program specific identifiers to another single special token. 

We define a code snippet to be an ordered union $C = \bar{U}(\Lambda, \Gamma, \mathbb{I})$. For all $t_{i} \in C$, we define an abstract structure representation of $C$ as:

\[
    \Tilde{C}= 
\begin{cases}
    t_{i},& \text{if } t_{i} \in \Lambda \\
    \phi(t_{i}),              & \text{if } t_{i} \in \Gamma \\
    \psi(t_{i}), & \text{if } t_{i} \in \mathbb{I}
\end{cases}
\]

We note that our strategy may not be the most optimal one in tasks like code token prediction or code completion, but it can be applied in situations where a representation of structure of code is required. Keeping this requirement in mind, we can justify mapping user defined variables/identifiers to special tokens as a way to reduce unwanted noise in input representation. In our experiments, we found that this representation qualitatively improves retrieved matches for a given COBOL snippet.

One of the key design choices we made in designing the abstract structural representation was to preserve as much \emph{context} as possible in a given code. Since code is a highly structured form of natural language~\cite{hindle_naturalness_nodate}, a transformation that preserves the ordering of tokens will produce a much richer representation. In contrast, representing a program as an unordered bag of root to leaf paths (in AST representation)~\cite{alon_code2vec_2019} destroys the ordering amongst different statements in a given program and can lead to different predictions if any statement is changed (while keeping the overall program correctness intact). We observe that our representation can extend beyond COBOL and can be applied to any programming language. 


\subsection{Unsupervised Learning Model}
\begin{figure}
    \centering
    \includegraphics[width=\textwidth]{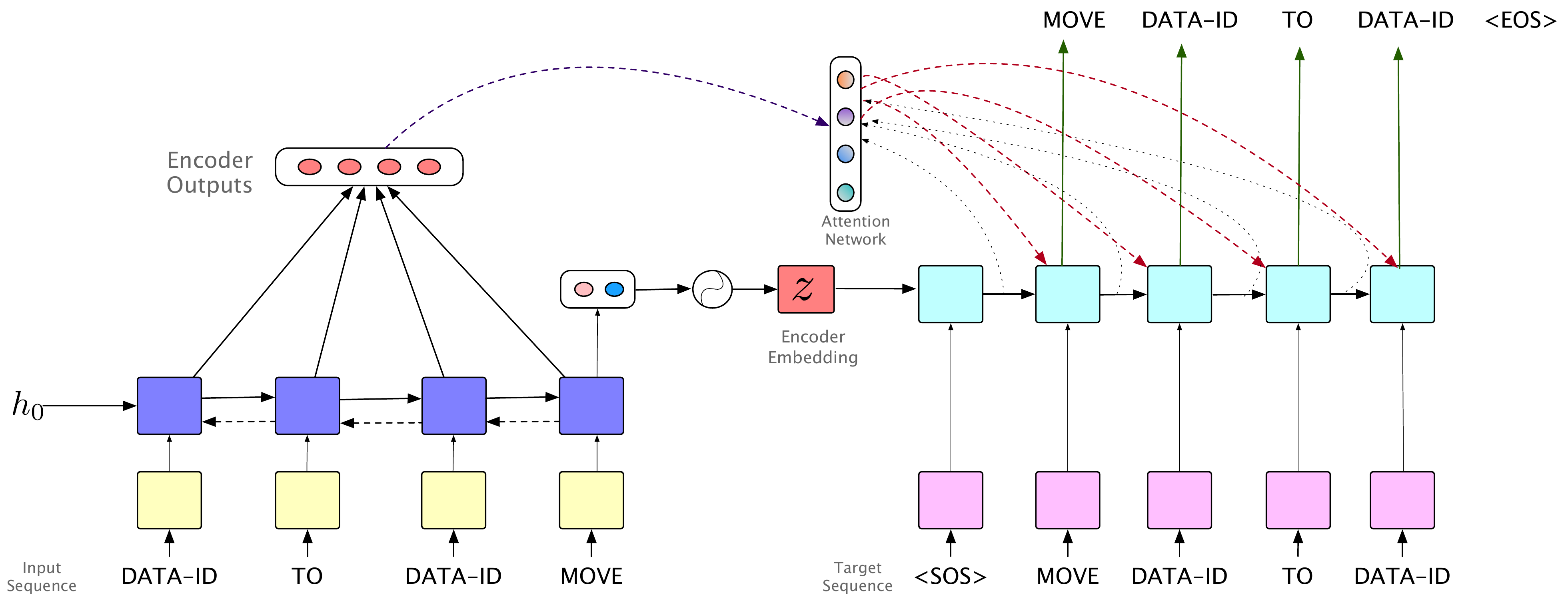}
    \caption{Seq2Seq Autoencoder for learning representation of code.}
    \label{fig:seq2seq_model}
\end{figure}

The abstract structure representation of code enables us to model lines of code as a sequence of tokens with an inherent ordering in between them. These sequences of tokens can be of arbitrary length. We wish to model the overall dependency between different tokens and the overall program structure in an unsupervised manner. This problem can be approached by viewing it as a sequence translation task with missing target sequences. In order to solve this problem we adapt the basic \texttt{seq2seq}~\cite{sutskever_sequence_2014} architecture and modify it to \emph{reconstruct} the original input sequence and in the process learn a fixed low dimensional representation of the sentence. We call this model a \emph{seq2seq autoencoder}. We decided on modifying seq2seq instead of using BERT-like architecture due to our limited data which would lead to the risk of over fitting the noise in the training set. Figure~\ref{fig:seq2seq_model} shows our overall architecture.

Let $\Tilde{C} = \{ x_{1}, x_{2} \dots x_{T} \}$ be the input code-token sequence of length T in the abstract structure representation. We obtain the fixed low dimensional vector $z$ as the last hidden state  of the  encoder.  At each step in decoding, we wish to learn the following conditional probability:

\begin{equation}
    p(\hat{x}_{1}, \hat{x}_{2} \dots \hat{x}_{T} | x_{1}, x_{2} \dots x_{T}) = \prod^{T}_{t=1} p(\hat{x}_{t} | z, x_{T}, x_{T-1} \dots x_{1})
    \label{eq:cond_prob}
\end{equation}

We append an end of sequence token \texttt{EOS} and initialize the decoding with a special start of sequence token (\texttt{SOS}) as well. For our loss function we chose to minimize the negative log likelihood of the probability in equation~\ref{eq:cond_prob}. Following~\cite{sutskever_sequence_2014} we also reversed the order of the input sequence and observed better convergence from our network. Code token sequences unlike natural language tend to have both backward and forward dependencies e.g.  in a statement like \texttt{MOVE X TO Y} there is a forward dependency between token \texttt{X} and token \texttt{Y} since a source register is moving it's value into a target register. At the same time, token \texttt{Y} also has a backward dependency on \texttt{X}. In order to capture the forward and backward dependencies we replaced the vanilla LSTM in the encoder with a bi-directional LSTM~\cite{biLSTM}. We compute $z = W. \text{tanh}([h^{(f)}_{T}; h^{(b)}_{T}])$ where $h^{(f)}_{T}$ and $h^{(b)}_{T}$ are the last hidden forward and backward states of the encoder respectively. We train our model to minimize the negative log likelihood between reconstructed tokens and the original source code tokens.

We also found that adding an attention module that computes the attention over source tokens with respect to current decoder state to improve model performance. We will explore the effects of attention in the coming sections.


\section{COBOL2Vec Dataset}

\begin{figure}
    \centering
    \subfigure[]{\includegraphics[width=.65\columnwidth]{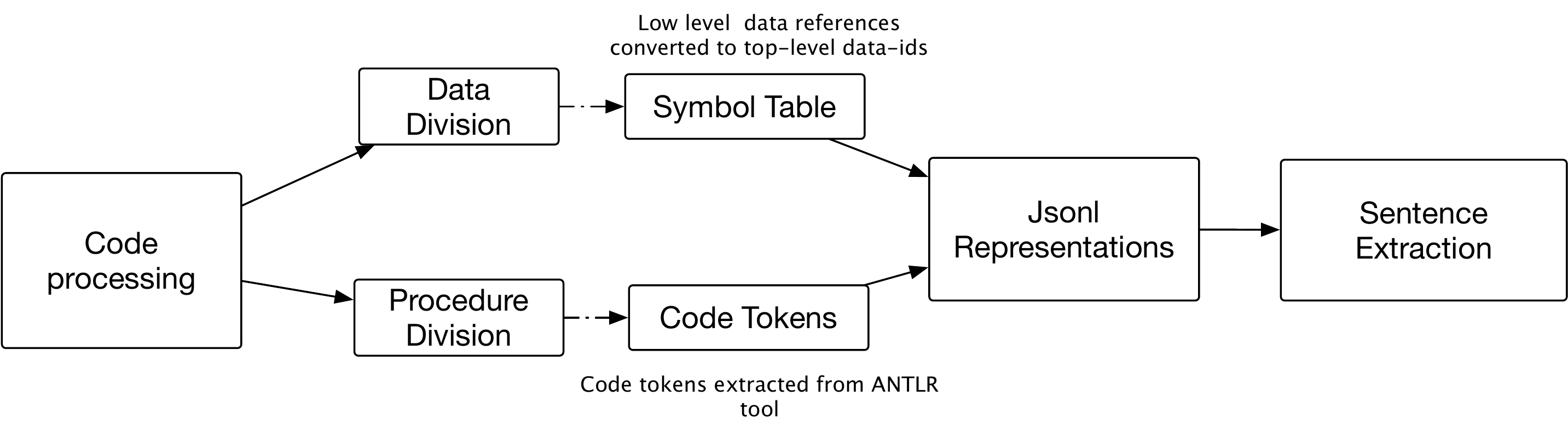}}
    \subfigure[]{\includegraphics[width=.65\columnwidth]{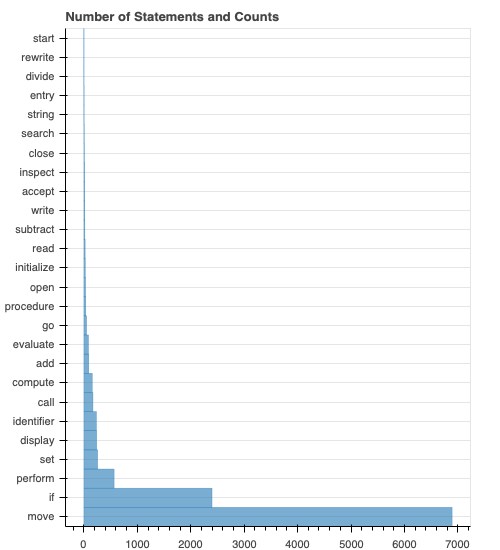}}
    \caption{Different parts of the Cobol2Vec Pipeline. (a) Overall pipeline schematic (b) Statistics of different type of statements in the corpus.}
    \label{fig:sstats}
\end{figure}

\begin{table}[h]
    \centering
    \begin{tabular}{c|c|c|c}
    \hline
        Sentence Type & Max Length & Min Length & Avg. Length \\
    \hline
    \hline
        PROCEDURE & $48$ & $4$ & $8.371$ \\
        PERFORM & $296$ & $4$ & $35.89$ \\ 
        IF & $298$ & $4$ & $45.37$ \\ 
        INITIALIZE & $20$ & $4$ & $7.64$ \\
        ACCEPT & $5$ & $4$ & $4.06$ \\ 
        MOVE & $290$ & $4$ & $9.15$\\ 
        OPEN & $167$ & $3$ & $23.42$ \\ 
        STRING & $30$ & $13$ & $19.0$ \\ 
        SET & $119$ & $4$ & $6.57$ \\ 
        CLOSE & $42$ & $4$ & $10.28$ \\ 
        ADD & $52$ & $4$ &  $6.92$ \\
        CALL & $70$ & $4$ & $12.917$\\
        IDENTIFIER & $134$ & $3$ & $11.76$ \\
        EVALUATE & $287$ & $23$ & $87.94$ \\
        READ & $84$ & $8$ & $18.78$ \\ 
        WRITE & $19$ & $4$ & $5.18$ \\ 
        GO & $3$ & $3$ & $3.0$ \\ 
        COMPUTE & $160$ & $4$ & $10.43$\\
        ENTRY & $48$ & $12$ & $36.0$\\
        SEARCH & $86$ & $25$ & $42.4$\\
        SUBTRACT & $10$ & $4$ & $4.35$\\
        DIVIDE & $10$ & $10$ & $10.0$\\
        INSPECT & $41$ & $8$ & $10.84$\\
        REWRITE & $20$ & $20$ & $20.0$\\
        START & $8$  & $8$ & $8.0$\\
    \end{tabular}
    \caption{COBOL2Vec sentence types and their lengths.}
    \label{tab:c2v_stats}
\end{table}

The lack of any open source COBOL repositories motivated our development of the \texttt{Cobol2Vec} dataset. We internally sourced COBOL files and processed them using the pipeline shown in Fig~\ref{fig:sstats}.

Processing COBOL code is different from high level languages like Java, Python in the way data and commands are represented in a program. In COBOL, all the data entities need to be defined in the \texttt{DATA-DIVISION} section of a COBOL program followed by the \texttt{PROCEDURE-DIVISION} for the code. While COBOL has code-books which allows for some separation between data and implementation, the code-books themselves need to be imported into the source file for proper compilation. Additionally, the variables can be nested in a hierarchical fashion. In order to keep the number of varying tokens as low as possible and convert to our abstract structure representation, we wanted to convert each data-identifier into a single token mapping.

We processed all source code files into an intermediate representation(IR) that processed all code-tokens,  keywords and their types in a single data structure. If any token was found to be a data-identifier then we computed the top-level variable name using an on-the-fly mapping discovered during the processing of the source file. This top level variable name is then converted to a single fixed token name to preserve structure. We informally call our lookup algorithm as a ``back-reference" lookup algorithm. 

We view the source code as a collection of small blocks of statements that perform a single action. In higher level languages, these can be constructs like functions, classes etc. In COBOL, the smallest construct is a \emph{sentence}. We extract sentences from the IR and convert them to the abstract structure representation. In order to achieve meaningful results we filter out sentences with lengths smaller than $3$ and greater than $300$.  Our resulting corpus has 11000 sentences sourced from internal COBOL files. However, the processing pipeline is general enough to work for other languages and only requires the information about the smallest granularity of source code unit. Figure~\ref{fig:sstats} shows the occurrences of different types of sentences in the corpus and  Table~\ref{tab:c2v_stats} shows the statistics about the different sentence types in the corpus.

\section{Experiments}

We trained our \texttt{seq2seq} autoencoder model on the Cobol2vec corpus for $50,000$ iterations on a single NVIDIA V-100 GPU. For all our experiments we used the Adam optimizer with a learning rate of $0.001$.  We used the traditional 80/20 train/test split and empirically chose an embedding size of $512$. We initialized the weights of all layers by drawing samples from a normal distribution $\mathcal{N}(0, 1)$. We chose to minimize the cross entropy loss between the source and target sequences and as a data augmentation we choose between using the next token in the target sequence as input and using the most likely token produced by the previous timestep with a probability of $0.5$. We found that this stabilized the training loss and produced much more accurate results.

\subsection{Discovering Structure in Code Embeddings}

A parallel and important direction in our work is to discover structure inherent in the overall code embeddings. This structural information can then help developers to make decisions about prioritizing certain closely linked code modules over others when an old code-base is ported to a newer language.

\begin{figure}[ht]
    \centering
    \includegraphics[width=.5\columnwidth]{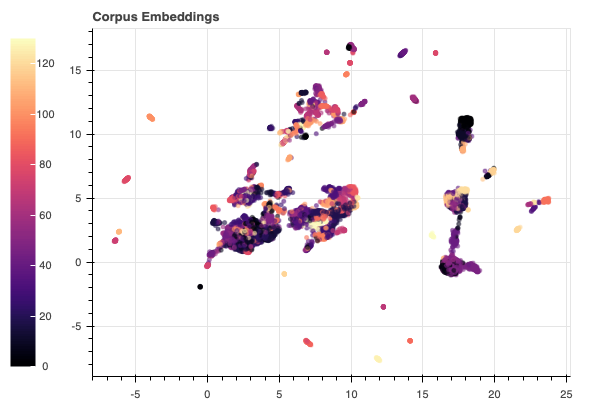}
    \caption{UMAP projection of Cobol2Vec corpus}
    \label{fig:c2v_umap}
\end{figure}

\begin{figure}[htbp]
    \centering
    \includegraphics[height=.5\columnwidth, width=.55\columnwidth]{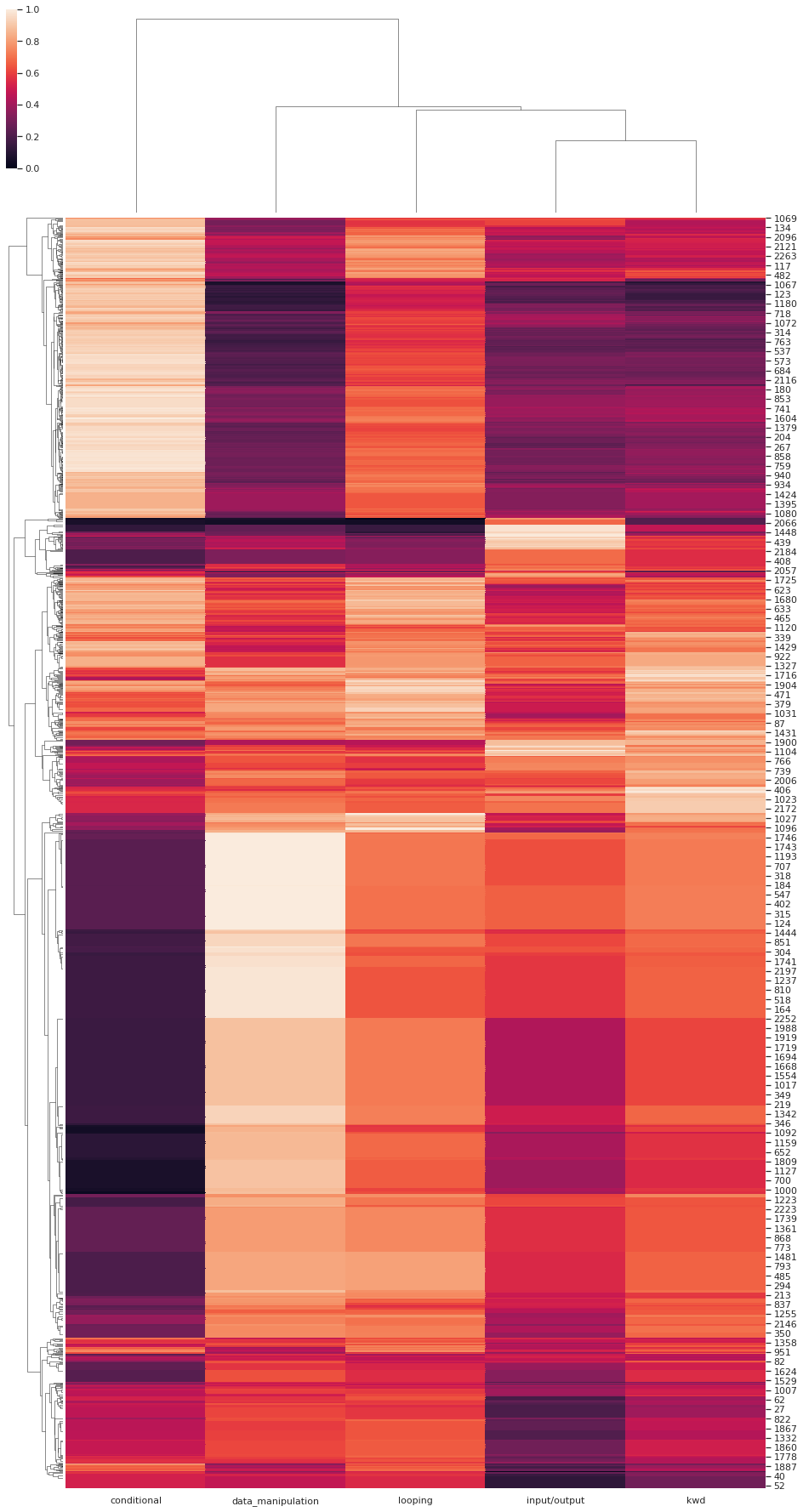}
    \caption{Agglomerative clustering across major types of statements in Cobol2Vec corpus.}
    \label{fig:c2v_cluster}
\end{figure}

In order to visualize the clusters of statements that belong closely together, we extracted the vector representations of all statements in the Cobol2Vec corpus by running our trained model on them. Instead of using t-SNE projection which does not always yield a good projection, we used UMAP~\cite{umap} to project the vectors in 2-d space. This projection is shown in Figure~\ref{fig:c2v_umap}. At a glance, this shows us a ``map" of our corpus. To further facilitate the interpretation of this plot, we built an interactive plot that allows the user to identify which statements belong to what cluster by simply hovering anywhere in the plot. This tool paired with our projection is a powerful way for any user to identify parts of their code that are closely linked together. Furthermore, the information gained from our projection can also help identify ``problematic" statements that may have vulnerabilties.

To further assist developers in identifying areas to focus on, we broadly categorized the cobol2vec corpus into 5 categories - \texttt{conditional}, \texttt{data\_manipulation}, \texttt{looping}, \texttt{i/o}, \texttt{kwds}. Where \texttt{kwd} is for reserved keywords. We performed an agglomerative clustering on the pre-existing vectors and constructed a heatmap that depicts the correlation of different type of statements with each other. The heatmap is shown in Figure~\ref{fig:c2v_cluster} and shows a corpus level view of all-pair similarity between different statements in the corpus. The dendrograms on the top and left of the heatmap show the coarse-grained and fine-grained clusters discovered via average linking. In the case of Cobol2Vec corpus, the major clusters are \texttt{conditional} and the rest of the types form other cluster. Within that cluster, \texttt{data\_manipulation} forms one component and the \texttt{looping} and \texttt{kwd} form others. We believe that this kind of clustering is most helpful when the code corpus size is small.





\subsection{Interpreting Attention}

\begin{figure}[h]
    \centering
    \includegraphics[width=\textwidth]{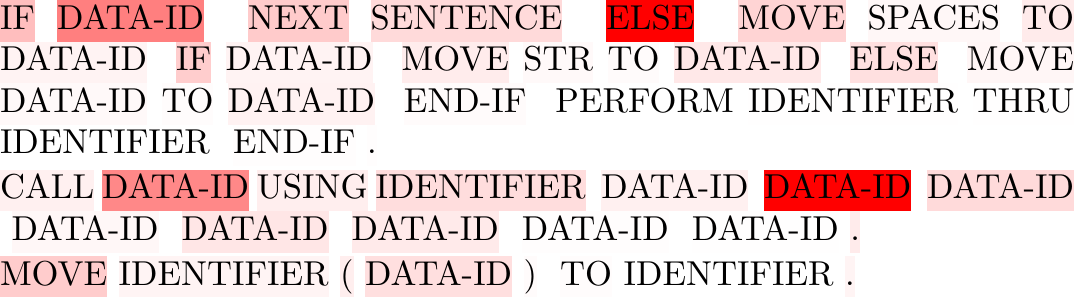}
    \caption{Attention on different tokens in code snippets.}
    \label{fig:att_sentence}
\end{figure}

In our experiments we noticed a significant improvement in the performance of the network when we added an attention mechanism to our decoder. While it's intuitive to understand how attention mechanism can help, we wanted to understand how precisely it helps the decoder make it's output decisions. Figure~\ref{fig:att_sentence}
shows how much weight is given to each token during decoding. A darker color indicates that a higher weight is placed on the token. 

In the first and third snippets, a higher weight is placed on the conditionals and the verb token respectively which makes sense since both these snippets are directing the compiler to take different actions based on the verb. However, in the second snippet the verb token gets a lower attention weight since it's more important to decide \emph{what} to call and \emph{where} to store the result of the value. Our observations clearly show that \emph{attention helps the decoder to focus more on the tokens based on the context}.

\section{Conclusion}

In this paper we have demonstrated an approach for learning a unique representation of old mainframe languages like COBOL. We designed an in house dataset and demonstrated that our hypothesis of treating code as a sequence of tokens rather than a bag of root to leaf paths is a valid one and showed that we can discover structure and retrieve closely related samples with our neural network. 

This direction of research opens up interesting avenues for future work. First, it presents an opportunity to build deep neural networks for learning a representation of old mainframe languages with a hugely different syntax than modern langauges. Second, it shows that code can be treated as a \emph{structured} natural language and thus \emph{context} plays a crucial role in deciding the performance of the algorithm. Thus, designing algorithms that engineer features from context is another interesting area of research.

\bibliographystyle{unsrt}
\bibliography{code_nlp}

\begin{thebibliography}{1}

\bibitem{mikolov2013efficient}
Tomas Mikolov, Kai Chen, Greg Corrado, and Jeffrey Dean.
\newblock Efficient estimation of word representations in vector space, 2013.

\bibitem{alon_code2vec_2019}
Uri Alon, Meital Zilberstein, Omer Levy, and Eran Yahav.
\newblock code2vec: learning distributed representations of code.
\newblock {\em Proc. ACM Program. Lang.}, 3(POPL):1--29, January 2019.

\bibitem{feng_codebert_2020}
Zhangyin Feng, Daya Guo, Duyu Tang, Nan Duan, Xiaocheng Feng, Ming Gong, Linjun
  Shou, Bing Qin, Ting Liu, Daxin Jiang, and Ming Zhou.
\newblock {CodeBERT}: {A} {Pre}-{Trained} {Model} for {Programming} and
  {Natural} {Languages}.
\newblock {\em arXiv:2002.08155 [cs]}, April 2020.
\newblock arXiv: 2002.08155.

\bibitem{lachaux_unsupervised_2020}
Marie-Anne Lachaux, Baptiste Roziere, Lowik Chanussot, and Guillaume Lample.
\newblock Unsupervised {Translation} of {Programming} {Languages}.
\newblock {\em arXiv:2006.03511 [cs]}, August 2020.
\newblock arXiv: 2006.03511.

\bibitem{hindle_naturalness_nodate}
Abram Hindle, Earl Barr, Mark Gabel, Zhendong Su, and Prem Devanbu.
\newblock On the {Naturalness} of {Software}.
\newblock page~12.

\bibitem{allamanis_mining_2013}
Miltiadis Allamanis and Charles Sutton.
\newblock Mining source code repositories at massive scale using language
  modeling.
\newblock In {\em 2013 10th {Working} {Conference} on {Mining} {Software}
  {Repositories} ({MSR})}, pages 207--216, San Francisco, CA, USA, May 2013.
  IEEE.

\bibitem{tufano_deep_2018}
Michele Tufano, Cody Watson, Gabriele Bavota, Massimiliano Di~Penta, Martin
  White, and Denys Poshyvanyk.
\newblock Deep learning similarities from different representations of source
  code.
\newblock In {\em Proceedings of the 15th {International} {Conference} on
  {Mining} {Software} {Repositories} - {MSR} '18}, pages 542--553, Gothenburg,
  Sweden, 2018. ACM Press.

\bibitem{devlin2019bert}
Jacob Devlin, Ming-Wei Chang, Kenton Lee, and Kristina Toutanova.
\newblock Bert: Pre-training of deep bidirectional transformers for language
  understanding, 2019.

\bibitem{sutskever_sequence_2014}
Ilya Sutskever, Oriol Vinyals, and Quoc~V Le.
\newblock Sequence to {Sequence} {Learning} with {Neural} {Networks}.
\newblock In Z.~Ghahramani, M.~Welling, C.~Cortes, N.~D. Lawrence, and K.~Q.
  Weinberger, editors, {\em Advances in {Neural} {Information} {Processing}
  {Systems} 27}, pages 3104--3112. Curran Associates, Inc., 2014.

\end{thebibliography}

\end{document}